\documentclass[twocolumn,aps,prb,groupedaddress]{revtex4-1}
\usepackage{graphicx}
\usepackage{color}
\newcommand\CBSOB{Cu$_3$Bi(SeO$_3$)$_2$O$_2$Br}
\newcommand{\half}{{\ensuremath{\frac{1}{2}}}}
\newcommand{\quat}{{\ensuremath{\frac{1}{4}}}}
\definecolor{red}{rgb}{1,0,0}
 
\definecolor{green}{rgb}{0,1,0}
 
\definecolor{blue}{rgb}{0,0,1}

\begin{document}

\title{Magnetic ground state and 2D behavior in pseudo-Kagom\'{e} layered system \CBSOB}

\author{M. Pregelj}
\affiliation{Josef Stefan Institute, Ljubljana, Slovenia}
\affiliation{Laboratory for Neutron Scattering, Paul Scherrer Insitute, CH-5232  Villigen, Switzerland}

\author{O. Zaharko}
\affiliation{Laboratory for Neutron Scattering, Paul Scherrer Insitute, CH-5232  Villigen, Switzerland }

\author{A. G\"unther}
\affiliation{Experimental Physics V, Center for Electronic
Correlations and Magnetism, University of Augsburg, 86135
Augsburg, Germany}

\author{A. Loidl}
\affiliation{Experimental Physics V, Center for Electronic
Correlations and Magnetism, University of Augsburg, 86135
Augsburg, Germany}

\author{V. Tsurkan}
\affiliation{Experimental Physics V, Center for Electronic
Correlations and Magnetism, University of Augsburg, 86135
Augsburg, Germany}
\affiliation{Institute of Applied Physics, Academy of Sciences of Moldova, MD-2028 Chisinau, Republic of Moldova}

\author{S. Guerrero}
\affiliation{Condensed Matter Theory, Paul Scherrer Insitute, CH-5232  Villigen, Switzerland }

\date{\today}

\begin{abstract}
Anisotropic magnetic properties of a layered Kagom\'{e}-like system {\CBSOB} have been studied by bulk magnetization and magnetic susceptibility measurements as well as powder and single crystal neutron diffraction. At $T_N$\,=\,27.4\,K the system develops an alternating antiferromagnetic order of ($ab$) layers, which individually exhibit canted ferrimagnetic moment arrangement, resulting from the competing ferro- and antiferro-magnetic intralayer exchange interactions.
A magnetic field $B_C\,\sim\,0.8$\,T applied along the $c$-axis (perpendicular to the layers) triggers a metamagnetic transition, when every second layer flips, i.e., resulting in a ferrimagnetic structure. Significantly higher fields are required to rotate the ferromagnetic component towards the $b$-axis ($\sim\,7$\,T) or towards the $a$-axis ($\sim\,15$\,T). The estimates of the exchange coupling constants and features indicative of a XY character of this quasi 2D system are presented.
\end{abstract}

\pacs{n/a}
\keywords{n/a} \maketitle

\section{INTRODUCTION}

The interest for novel frustrated layered compounds stems from their compelling magnetic properties, which attract applied as well as basic science oriented research.
In case of a strong ferromagnetic (FM) intralayer exchange and a weak antiferromagnetic (AFM) interlayer coupling, the magnetic ground state of AFM arranged FM layers can 
be easily broken by external magnetic fields, which enable simple switching between zero and maximum magnetization. 
The apparent metamagnetic \cite{Jacobs} response -- an abrupt change of the bulk magnetization -- is thus most appreciated in high-density magnetic storage and spintronics devices. \cite{Wolf}
Such systems are often described by models on the two-dimensional (2D) lattice with spin dimensionality $n$\,=\,1 (Ising), $n$\,=\,2 (XY) or $n$\,=\,3  (Heisenberg),\cite{Jongh90} which are interesting also from the theoretical point of view. The 2D XY model exhibits a unique feature -- a Kosterlitz-Thouless transition \cite{KT} from a paramagnetic state to a phase with quasi long-range spin order with vortex and antivortex exitations.
On contrary, the 2D Ising model shows conventional long-range order, \cite{Onzager}
whereas the 2D Heisenberg model does not order at any finite temperature. \cite{Mermin}
Yet, real materials are usually more complex.
Frequently the planar rotational symmetry is imperfect, leading to a quasi XY behavior. The inter-plane exchange coupling is often sufficiently strong to induce 3D ordering, so also spatially such systems are only quasi 2D. Finally, layered systems, particularly frustrated ones, \cite{Lacroix} are highly susceptible to small external perturbations \cite{Landau81} allowing to sweep across spin- and space- degrees of freedom.\cite{Pelissetto} 

\begin{figure}[!]
\includegraphics[width=90mm,height=100mm,keepaspectratio=true,angle=0,trim=0mm 10mm 0mm 10mm]{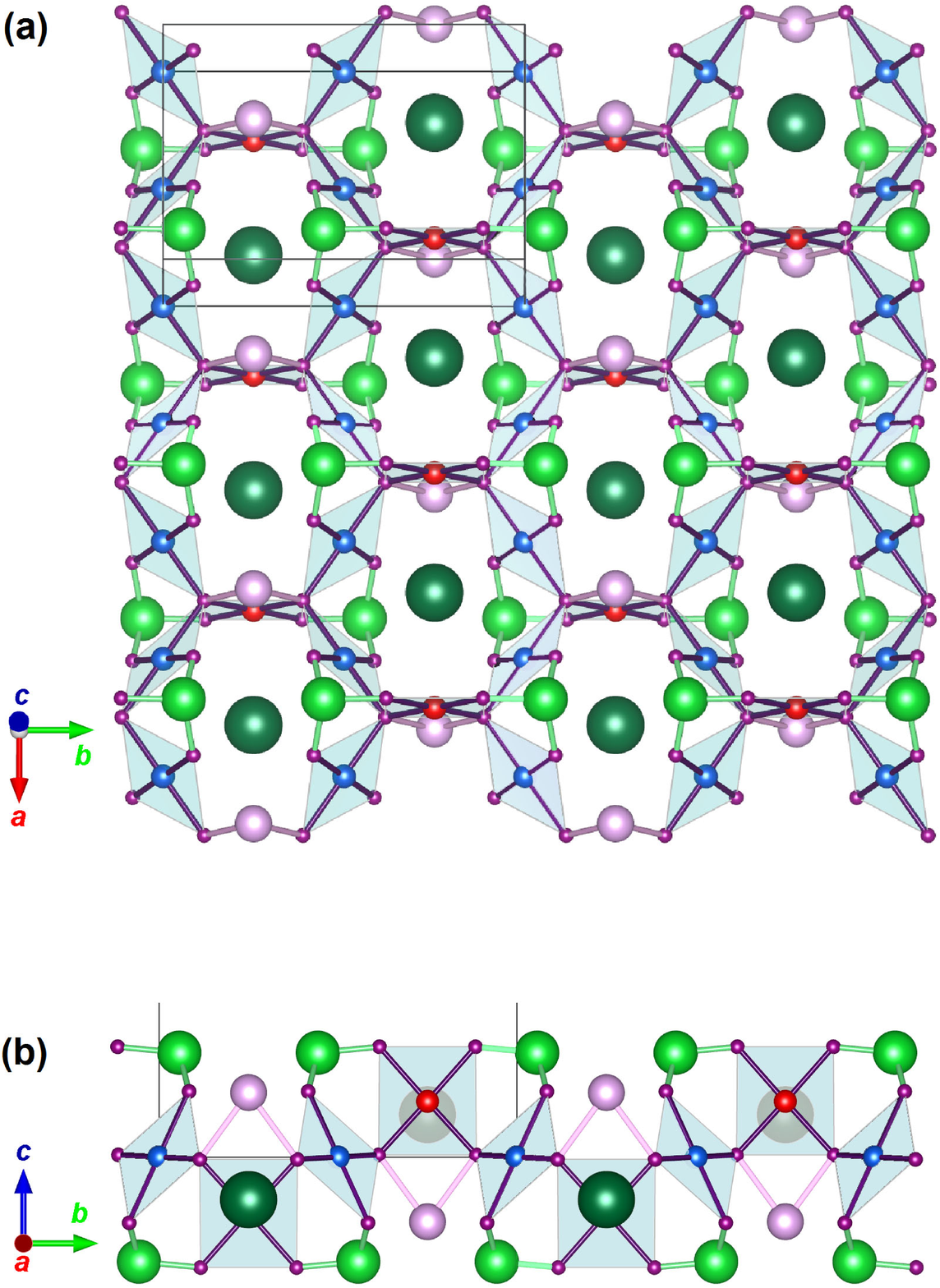}
\caption {(Color online) (a) $ab$ and (b) $bc$ projections of the {\CBSOB} crystal layer. For clarity the $ab$ projection is slightly canted. Here the magnetic Cu1 and Cu2 ions are indicated as 
small blue and red spheres, respectively; O ions are the smallest violet spheres in the corners of CuO$_4$ plackets, Bi ions are intermediate light pink spheres, Se ions are larger light green spheres, and Br ions are the largest dark green spheres.}
\label{figCryst}
\end{figure}

A novel layered compound with seemingly frustrated magnetic lattice is {\CBSOB}.
This compound is orthorhombic (space group $Pmmn$) with crystal lattice parameters $a$\,=\,6.390\,\AA, $b$\,=\,9.694\,\AA\, and $c$\,=\,7.287\,\AA. \cite{Millet01}
It is built of two different types of [CuO$_4$] square plackets, sharing apices to form copper-oxygen layers reminiscent of a buckled Kagom\'{e} lattice of the magnetic Cu$^{2+}$ ($S$\,=\,$\half$) ions (Fig.\,\ref{figCryst}), with Cu1 and Cu2 positioned at (000) and (\quat \quat $z$, $z$\,=\,0.791), i.e., at the 4(c) and the 2(a) sites, respectively.
Hence, in case of the AFM nearest-neighbor interactions, Cu$^{2+}$ spins
 should be exposed to a strong geometrical frustration. However, additional Cu-O-$X$-O-Cu
 ($X$\,=\,Se, Bi) super-superexchange interactions might be important,\cite{Deisenhofer} as the
 [CuO$_4$] plackets are linked  also by Se$^{4+}$ and Bi$^{3+}$ ions 
(Fig.\,\ref{figCryst}). In fact, Bi$^{3+}$ enables 
 additional Cu1-O-Bi-O-Cu1 next-nearest-neighbor interaction along the $b$ axis, whereas Se$^{4+}$ provides only an alternative to the already recognized Cu-O-Cu nearest-neighbor interactions.
 Finally, since both, Se$^{4+}$ and Bi$^{3+}$, possess lone pair electrons and thus effectively reduce the number of chemical bonds, \cite{Johnsson} the only probable interlayer coupling is through a long Bi$^{3+}$-O bond.

Earlier powder magnetic susceptibility, \cite{Millet01} measured at 1\,T, suggests dominant FM interactions, as its high-temperature behavior above 150\,K follows a Curie-Weiss (CW) law $\chi\,=\,C/(T-\theta)$ with a FM Weiss temperature $\theta\approx$\,57\,K. At lower temperatures, a weak anomaly, associated with tiny structural changes, \cite{Millet01} occurs at $\sim$\,120\,K, whereas a sharp step, implying establishment of long-range magnetic order, is found at $T_N\approx$\,24\,K.

Here we present a detailed magnetic susceptibility, magnetization and neutron diffraction study of the low-temperature magnetic ground states and metamagnetic transition in {\CBSOB}. 
We find that below $T_N$ individual ($ab$) layers with a canted ferrimagnetic spin arrangement order antiferromagnetically. When $B_C$\,$\sim$\,0.8\,T is applied perpendicular to the layers (along $c$), weak AFM interlayer interactions are suppressed and a metamagnetic transition is triggered, as every second layer flips --  resulting in an overall canted ferrimagnetic structure.
Based on the determined spin arrangements and critical fields of the metamagnetic transition we provide an estimate of the characteristic exchange couplings of the system. 
Interestingly, though the ratio between the interlayer and the effective intralayer couplings is significant,  $|J'/J|$\,$\sim$\,0.006,  and the magnetic anisotropy represents only $\sim$\,20\,\% of $J$, the low temperature experimental data sustain a quasi 2D XY behavior.

\section{EXPERIMENTAL DETAILS}

Polycrystalline material of {\CBSOB} was prepared by solid-state reactions at 550$^\circ$C from the high-purity binary compounds. Single crystals were grown at 500 - 550$^\circ$C by the chemical-transport-reaction method with bromine as the transport agent.  
Magnetization ($M$) and dc susceptibility ($\chi=M/H$) measurements were performed in commercial MPMS and PPMS magnetometers in the temperature range 1.8\,K - 400\,K and applied magnetic fields up to 8 T.\\

Neutron diffraction experiments were performed at Swiss Neutron Spallation Source SINQ, Paul Scherrer Institute, Switzerland. 
Powder diffraction patterns were collected between 1.5\,K and 60\,K on the DMC powder diffractometer with neutron wavelength $\lambda$ = 2.46 \AA. Single crystal diffraction was performed in the temperature range between 6.5\,K and 300\,K on the TriCS single crystal diffractometer ($\lambda$ = 1.18 \AA). The zero-field data collection was performed in a cooling machine mounted on a 4-circle cradle, for the magnetic field measurement a vertical field cryomagnet (Oxford Instruments) and normal beam geometry were used.

\section{RESULTS}

\subsection{Magnetization and magnetic susceptibility}

\begin{figure}[!]
\includegraphics[clip,width=90mm,keepaspectratio=true,angle=0,trim=0mm 0mm 0mm 10mm]{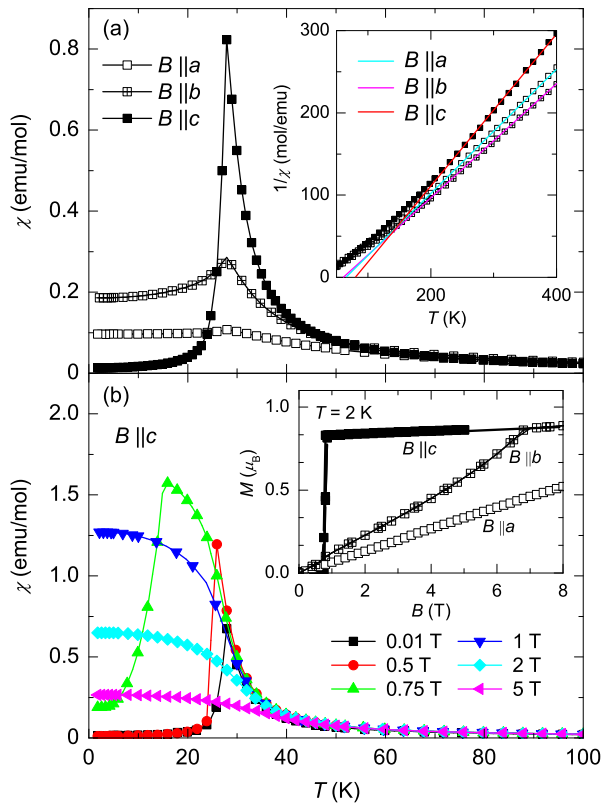}
\caption {(Color online) (a) Magnetic susceptibility at 0.01 T along all three crystallographic axes.  Inset: the inverse susceptibilities (symbols) and fits (solid lines) to the Curie-Weiss laws. (b) Magnetic susceptibility measured in different external magnetic fields $B||c$ between 0.01 - 5 T. Inset: magnetization at 2\,K measured along all three crystallographic directions.}
\label{figChiBpaC}
\end{figure}

Compared to earlier reported powder magnetic susceptibility data measured in a magnetic field of 1 T, \cite{Millet01} our single crystal results reveal new anisotropic properties of {\CBSOB}. In fact, we find that the magnetic susceptibility is very anisotropic and highly sensitive to the strength of the applied magnetic field. 
When an external magnetic field of 0.01\,T is applied perpendicular to the layers ($B||c$), $\chi_c(T)$ can be described by a CW law with $\theta_c$\,=\,80(5)\,K down to 200\,K, where the slope of 1/$\chi_c(T)$ starts to decrease (inset in Fig.\,\ref{figChiBpaC}a). 
For perpendicular orientations, i.e., the field applied within the layers ($B || a,b$), the high temperature behavior of $\chi_{a,b}(T)$ shows similar response (Fig.\,\ref{figChiBpaC}a), with $\theta_{a,b}$\,=\,60(5)\,K.
We note that the difference in 1/$\chi_c(T)$ and 1/$\chi_{a,b}(T)$ slopes suggests different $g$-factors, as Curie constant $C$\,$\propto$\,$g^2$,\cite{Kittel} while the discrepancy between $\theta_{c}$ and $\theta_{a,b}$  implies a sizable ($\sim$\,20\,\%) exchange anisotropy.\cite{Han12}

A much more anisotropic response is observed below the magnetic transition (for $T$\,$<$\,$T_N$\,=\,27.4\,K), where a typical AFM behavior is observed at low fields (Fig.\,\ref{figChiBpaC}a). For $B||c$ the anomaly in $\chi_c(T)$ is very sharp i.e., $\chi_c(T)/\chi_c(T_N)$ drops below 0.1 already at $T/T_N$\,$\sim$\,0.85, which implies that the magnetic ground state is AFM and that the $c$-axis  is the orientation favored by the magnetic moments.
Considering now that the anomalies in $\chi_{a}(T)$ and $\chi_{b}(T)$ are significantly less pronounced and that $\chi_a(T)$ is smaller than $\chi_b(T)$, we can deduce that the magnetic moments are more easily bent towards $b$ than towards $a$. Thus we can conclude that {\CBSOB} is an AFM with magnetic moments predominately lying along the $c$-axis, $a$ coincides with the magnetic hard axis and $b$ is somewhere in-between.

When the applied magnetic field $B||c$ exceeds 0.5\,T, the anomaly in $\chi_c(T)$ starts to broaden and shift to lower temperatures (Fig.\,\ref{figChiBpaC}b). Eventually, at 1\,T, $\chi_c(T)$ exhibits saturation-like behavior, indicating that the low-field (LF) AFM ground state is followed by a high-field (HF) ferri- or ferromagnetic phase.
On contrary, increase of $B$\,$\perp$\,$c$ up to 5 T has almost no effect on $\chi_{a,b}(T$$<$$T_N)$ (not shown).

Finally, low-temperature (at 2\,K) magnetization $M$ was measured along all three crystallographic axes (inset in Fig.\,\ref{figChiBpaC}b). 
A sharp magnetic transition is found for $B_C$\,=\,0.8\,T with $B||c$. The fact that above $B_C$ the value of $M$ per Cu$^{2+}$ ion, $\mu_{Cu}$\,=\,0.9\,$\mu_B$, is almost constant and very close to 1\,$\mu_B$, expected for the complete magnetization per Cu$^{2+}$ ($S$\,=\,\half) ion, indicates that we witness a metamagnetic transition from an AFM to a ferro/ferrimagnetic state.
A very weak field dependence of $M$ above $B_C$ indicates that fully FM state is not reached yet. This suggests that a small AFM component is still present and that significantly stronger magnetic fields are required to overcome the responsible AFM interactions.
On the other hand, when $B$ is applied along the $b$-axis, the magnetization linearly increases up to 7\,T, where it again reaches a saturation value. Similar response is observed also for $B||a$, where the slope of the magnetization is a bit smaller, implying that saturation is reached at $\sim$\,15\,T. 
Relatively slow linear response for $B\perp c$ compared to $B || c$ indicates the presence of the sizable magnetic anisotropy. Since the zero-field splitting, imposed by the crystal field, is expected to be negligible in the Cu$^{2+}$ ($S$\,=\,1/2) systems, \cite{Kittel} we suspect that anisotropic exchange interaction is responsible for the observed behavior -- in agreement with the different values of the CW constants (inset in Fig\,\ref{figChiBpaC}a).

\subsection{Neutron diffraction}

\subsubsection{Powder diffraction}

\begin{figure}[!]
\includegraphics[width=90mm,keepaspectratio=true,angle=0,trim=4mm 0mm 3mm 7mm]{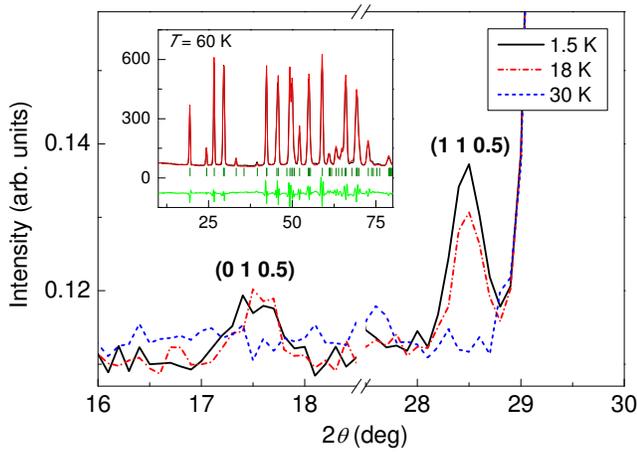}
\caption {(Color online) The enlarged part of the neutron powder diffraction pattern, in which the two most pronounced magnetic reflections appear, measured at several temperatures. Inset: The measured (black line) and the calculated (red line) powder neutron diffraction patterns at 60 K, whereas the green line shows the difference between the two, and the markers show the calculated positions of the Bragg reflections.}
\label{fig1}
\end{figure}

To determine the spin arrangement in the LF and HF states we first performed neutron powder diffraction measurements.
The diffraction pattern measured at 60\,K (inset in Fig.\,\ref{fig1}) confirms the crystal structure published by Millet et al., \cite{Millet01} while the patterns collected below 30 K reveal the onset of long-range magnetic order (Fig.\,\ref{fig1}). Magnetic reflections occur at positions distinct from the nuclear ones, indicating that the magnetic ground state is AFM. Indeed, the main magnetic reflections (2$\theta$\,=\,17.5\,$^\circ$ and 28.5\,$^\circ$) can be indexed by the magnetic wave vector {\bf k}\,=\,(0\,0\,\half). 

In the next step we performed the representation analysis of possible magnetic structures based on the propagation wave vector and crystal symmetry. \cite{FULLPROF} The little group consists of eight elements and the resulting irreducible representations (IRR) $\Gamma_i$, ($i$ = 1-8) are given in Table\,\ref{tab1}. It was, however, impossible to determine the unique magnetic ground state from the powder diffraction data alone. 

\begin{table} [!]
\caption{Irreducible representations for the Cu1 and Cu2 4(c) and 2(a) sites of the $Pmmn$ space group and the {\bf k}\,=\,(0\,0\,\half) wave vector.
\label{tab1}}
\begin{ruledtabular}
\begin{tabular}{ccccccc}
\multicolumn{2}{c}{site Cu1}& $\Gamma_1$ & $\Gamma_3$ & $\Gamma_5$ & \multicolumn{2}{c}{$\Gamma_7$}\\
\hline
\multicolumn{2} {c}{x,y,z          }&  u,v,w    &  u,v,w   &  u,v,w   &   \multicolumn{2}{c}{u,v,w} \\
\multicolumn{2} {c}{-x+$\frac{1}{2}$,-y+$\frac{1}{2}$,z  }& -u,-v,w   & -u,-v,w  &  u,v,-w  &   \multicolumn{2}{c}{u,v,-w} \\
\multicolumn{2} {c}{-x,y+$\frac{1}{2}$,-z      }& -u,v,-w   &  u,-v,w  & -u,v,-w  &   \multicolumn{2}{c}{u,-v,w}  \\
\multicolumn{2} {c}{x+$\frac{1}{2}$,-y,-z     }&  u,-v,-w  & -u,v,w   & -u,v,w   &   \multicolumn{2}{c}{u,-v,-w} \\
\hline
\hline
site Cu2& $\Gamma_2$ & $\Gamma_3$ & $\Gamma_5$ & $\Gamma_6$ & $\Gamma_7$ & $\Gamma_8$ \\
\hline
 x,y,z      & 0,0,u   & 0,0,u  & 0,u,0  &  u,0,0  & u,0,0  &  0,u,0  \\
-x,y+$\frac{1}{2}$,-z & 0,0,-u  & 0,0,u  & 0,u,0  & -u,0,0  & u,0,0  &  0,-u,0 \\
\end{tabular}
\end{ruledtabular}
\end{table}
\subsubsection{Single crystal diffraction}

\begin{figure}[!]
\includegraphics[width=160mm,height=85mm,keepaspectratio=true,angle=-90,trim=20mm 0mm 20mm 0mm]{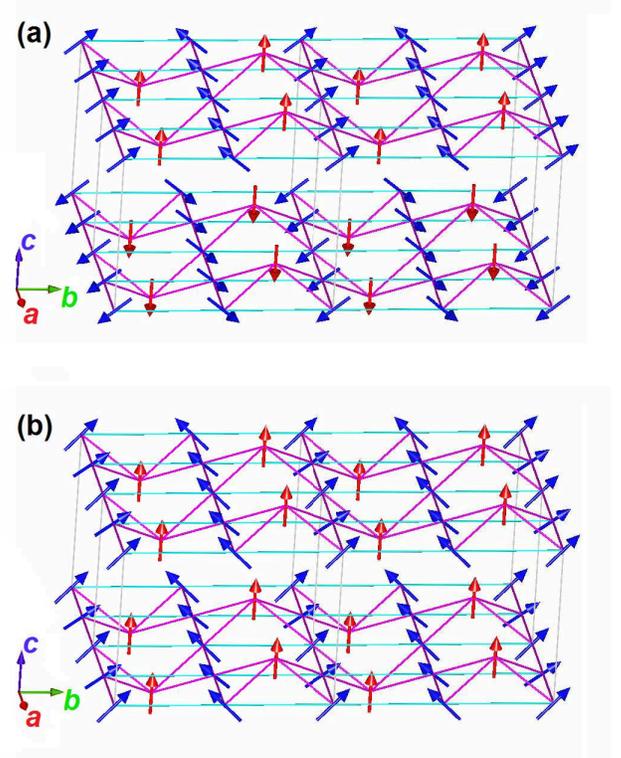}
\caption {(Color online) Refined magnetic structure at (a) zero applied field and (b) $B_C$=1\,T. Magnetic moments at the Cu1 sites are blue arrows, while Cu2 are red. For clarity only the magnetic lattice is show, i.e., $J_1 \approx J_3$ (violet) and $J_2$ (cyan) exchange interactions.}
\label{fig3}
\end{figure}

Single crystal diffraction was performed on a crystalline platelet with size $10\times10\times1$\,mm$^3$. To determine the LF magnetic ground state, 65 magnetic and 30 nuclear reflections were collected in zero-field at 6.5\,K, i.e., well below $T_N$. The best fit \cite{FULLPROF} was obtained for the magnetic structure described by the $\Gamma_3$ irreducible representation for both Cu sites. The refined magnetic components are $m_y=0.72(2)\,\mu_B$ and $m_z=0.57(2)\,\mu_B$ for Cu1 and $m_z=0.90(4)\,\mu_B$ for Cu2. This corresponds to the magnetic moment arrangement presented in Fig.\,\ref{fig3}a.
The Cu1 moments are aligned parallel to the $c$-axis with additional alternating component along the $b$-axis and are thus canted $\sim$\,$\pm$50\,$^\circ$ from $c$ towards $b$. The Cu2 moments, on the other hand, are strictly parallel to the $c$-axis.  
Such an arrangement suggests dominant FM interactions within the layer, as well as a sizable magnetic anisotropy. Yet the coupling between the layers is AFM as presented in Fig.\,\ref{fig3}a.

The temperature dependence of the (2\,2\,\half) magnetic reflection (Fig.\,\ref{fig4}a) indicates a continuous second order phase transition from the paramagnetic to the LF AFM phase. 
Below 26.6\,K, i.e. $1-T$/$T_N >$ 0.04, the magnetic intensity can be described by $I\sim(T_N-T)^{2\beta}$ with $\beta$\,$\leq$\,0.23 (Fig.\,\ref{fig4}d), as expected for 2D XY spin systems, \cite{Bramwell93, Bramwell94} in particular when additional weak in-plane crystal-field anisotropy is present.\cite{Taroni}
The exact $T_N$ was derived by fitting the data in the vicinity of $T_N$ (Fig.\,\ref{fig4}a), where the critical exponent $\beta$ increases to $\sim$\,0.30(2), signifying a crossover from spatial 2D (n=2) to 3D (n=1) behavior. Our result is in accord with observations for other layered systems e.g., K$_2$CuF$_4$, BaNi$_2$(PO$_4$)$_2$ and Rb$_2$CrCl$_4$,\cite{Hirakawa, Jongh90, Regnault90, Bramwell95} and implies that the 2D critical region extends in the magnetic long-range ordered state as well as that spin fluctuations are essentially two dimensional. We presume that spin waves in this material will be significantly renormalized by vortex excitations.\cite{Kosterlitz74, Jose}
A typical feature of the 2D regime -- diffuse magnetic scattering -- has yet not been detected, presumably due to smallness of the single crystal used in the diffraction experiment.

\subsubsection{Single crystal diffraction in applied magnetic field $B || c$}

In order to explore the HF state, i.e., above $B_C$\,=\,0.8\,T ($B$\,$\parallel$\,$c$), we measured the field dependence of the (2\,2\,\half) and (2\,1\,0) magnetic reflections at 1.5\,K (Fig.\,\ref{fig5}). As anticipated, the (2\,2\,\half) reflection abruptly disappears at $\sim$\,0.8\,T. On contrary, the (2\,1\,0) \cite{LowSim} reflection exhibits exactly the opposite response, which implies that it corresponds to the FM HF phase. This sharp transition is reminiscent of the metamagnetic behavior, suggesting that the AFM coupling between the adjacent layers is broken.
\begin{figure}[!]
\includegraphics[width=90mm,height=180mm,keepaspectratio=true,angle=0,trim=0mm 0mm 0mm 10mm]{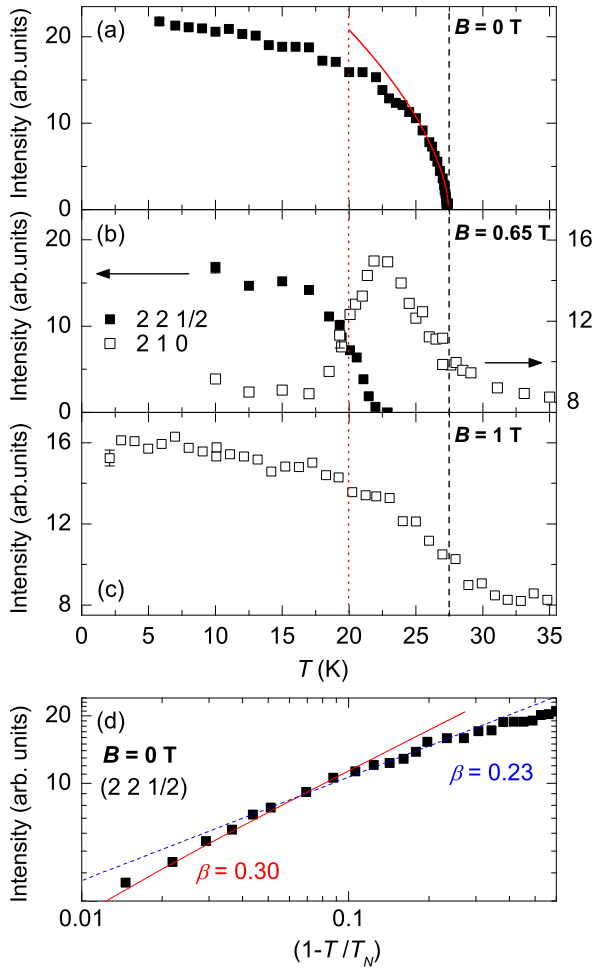}
\caption {Temperature dependences of the intensities of the (2\,2\,\half) and (2\,1\,0) reflections at (a) 0\,T, (b) 0.65\,T and (c) 1\,T. The black dashed line indicates the magnetic transition at $T_{N}$, while the red one indicates the temperature, at which field dependences were measured. The solid red lines represents a fit to $I\sim(T_N-T)^{2\beta}$, with $\beta$ =0.30(2) and $T_N=27.4$\,K. (d) Intensity of the (2\,2\,\half) magnetic reflection as a function of ($1-T/T_N$), where solid and dashed lines denote two distinct $I\sim(T_N-T)^{2\beta}$ regimes.}
\label{fig4}
\end{figure}
\begin{figure}[!]
\includegraphics[width=90mm,height=180mm,keepaspectratio=true,angle=0,trim=0mm 5mm 0mm 8mm]{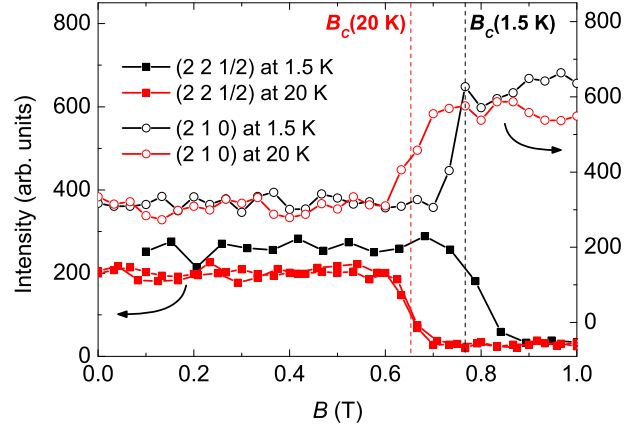}
\caption {(Color online) Field dependence of the intensities of the (2\,2\,\half) and (2\,1\,0) reflections at 1.5\,K and 20\,K.}
\label{fig5}
\end{figure}
To investigate the phase boundaries, we thus performed a series of temperature scans of the (2\,2\,\half) and (2\,1\,0) reflections at 0\,T, 0.65\,T and 1\,T (Fig.\,\ref{fig4}). The results corroborate with the magnetic susceptibility data, i.e., the applied magnetic field suppresses the LF phase and induces the HF phase. At 0.65\,T (Fig.\,\ref{fig4}b) the LF phase persists up to $\sim$\,19\,K, both phases coexist between 19\,K and 22\,K, implying a first-order metamagnetic transition, while between 22\,K and $T_N$ only the HF phase is left. At 1\,T, the LF phase is completely suppressed, while $T_N$\,=27.4\,K seems to be field independent. 

To determine the HF magnetic structure we collected 40 reflections, from which 17 had significant intensity. The best refinement was obtained for {\bf k}\,=\,(0\,0\,0) retaining the $\Gamma_3$ IRR for both Cu sites. 
The order within the layers appears to be almost unaffected, while the arrangement of the consecutive layers is now FM (Fig.\,\ref{fig3}b). 
At 1.5\,K and 1\,T for the Cu1 sites $m_x=-0.2(1)\,\mu_B$, $m_y=0.73(2)\,\mu_B$ and $m_z=0.66(4)\,\mu_B$ ($m=0.99(2)\,\mu_B$) and for the Cu2 sites $m_z=0.99(9)\,\mu_B$. 

\section{DISCUSSION}

The magnetic properties of {\CBSOB} presented in this work reflect the low dimensional (2D) nature of its lattice.
This is most evident from the experimentally determined magnetic structures for the LF and HF phases, which indicate that the main magnetic 'building block' of the system is a single layer. In order to understand the observed behavior, we thus focus here on the individual layer.
The two Cu sites have different local symmetries of their square planar [CuO$_4$] coordination, which might be responsible for the different behavior of Cu1 and Cu2 magnetic moments. 
However, calculation of the exchange charge model of the crystal field around Cu$^{2+}$ ions in the actual surrounding indicates that the zero-field spitting, imposed by the crystal field, is weak,\cite{Sophia} as expected for $S$\,=\,1/2 systems.
Another source of the observed response, e.g.,  magnetic frustration or exchange anisotropy, might lie within the exchange network. In particular, the electron hopping, governing the exchange interaction, appears to be very different for inter- and intra-layer exchange pathways and might also differ between the three exchange pathways within the layer (see Figs.\,\ref{figCryst} and \ref{fig3}). Two of these are nearest-neighbor ones, involving Cu1-O1-Cu1 ($d$\,=\,3.19\,\AA, bond angle $\phi$\,=\,111\,$^\circ$, multiplicity $m$\,=\,4) and Cu1-O1-Cu2 ($d$\,=\,3.27\,\AA, $\phi$\,=\,113\,$^\circ$, $m$\,=\,8) superexchange bridges, while the last connects next-nearest-neighbors via the Cu1-O1-Bi-O1-Cu1 ($d$\,=\,4.84\,\AA, $m$\,=\,4) super-superexchange bridge. In the above order we assign them the coupling constants $J_1$, $J_3$, and $J_2$, respectively. 
Since the local symmetry of the cations is low (point groups $S_2$ and $C_{2v}$) and crystalline fields arising from the surrounding oxygens deviate significantly from the usual octahedral and tetrahedral coordinations, reliable prediction of the sign and strength of the exchange interactions just by following the Goodenough-Kanamori-Anderson rules \cite{Goodenough63, Goodenough, Kanamori} is rather bold.
Nevertheless, the arrangements of the magnetic moments in the determined magnetic structures imply that the $J_1$ and $J_3$ couplings are most probably FM, while the $J_2$ exchange is AFM. 

In order to determine the above presented exchange interactions, we first explore all possible combinations of their sign and strength within the model of isotropic Heisenberg interactions, using the ENERMAG program.\cite{ENERMAG} Here, we assume that $J_1$ and $J_3$ are equal, which is justified by their similarity in bonding lengths and angles. As a result, we obtain a list of probable ordering modes for a specific magnetic propagation vector {\bf k} and corresponding intervals of $J_i$ ($i$\,=\,1,2).
We find that the only modes, which correspond to $\Gamma_3$ IRR on both Cu-sites (see Table\,\ref{tab1}) and thus meet the experimental observations, are M$_z$\,(+\,+\,+\,+\,+\,+) and M$_y$\,(+\,$-$\,$-$\,+\,0\,0). Furthermore, we discover that for {\bf k}\,=\,(0\,0\,$\half$) (LF phase) and {\bf k}\,=\,(0\,0\,0) (HF phase) the M$_z$\,(+\,+\,+\,+\,+\,+) mode is favorable, when $J_1$ is FM, and $J_2$ can have any sign; whereas the M$_y$\,(+\,$-$\,$-$\,+\,0\,0) mode is obtained, when all couplings are AFM. Apparently, these two modes require different sign of couplings and thus implies that the experimentally determined canted magnetic structure, which is a convolution of the two modes, is a result of competing magnetic interactions.

To resolve the canting of the Cu1 spins we thus minimize the energy of the magnetic ground state. In line with the experimental observations and to avoid unnecessary complications, we restricted the orientation of the Cu1 and Cu2 magnetic moments to the $bc$ plane. The results show that the experimentally determined canting of the Cu1 magnetic moments ($\sim$50\,$^\circ$ from $c$ towards $b$) can be achieved with isotropic, yet different, interactions. In fact, we find that $J_2$ should be AFM and even stronger than FM $J_1$, i.e., $J_2$\,$\sim$\,$-1.6J_1$. This seems contra-intuitive, as the exchange path for $J_2$ is significantly longer than those corresponding to $J_1$ and $J_3$, and since it involves an 
additional Bi$^{3+}$ ion. However, studies of other tellurides and selenides reveal that similar super-superexchange paths can have similar strengths as some significantly shorter superexchange bonds. \cite{Deisenhofer}

In order to quantify the strengths of the exchange interactions we take the Curie-Weiss
temperature into consideration, which is defined as the sum of the exchange interactions per magnetic site. Thus we can write the expression:
\begin{equation}
\theta_{CW} = \frac{2}{3 k_B} S(S+1) \sum_{n, i} ( J_i z_{ni}).
\end{equation}
Here, we assumed $(g-1)^2$\,$\approx$\,1, $n$\,=\,1,2 counts different Cu sites, $i$\,=\,1,2 counts different exchange interactions ($J_3$\,=\,$J_1$), and $z_{ni}$ is half of the number of Cu neighbors coupled by $J_i$ exchange ($z_{11}$=2, $z_{12}$=1, $z_{21}$=2, $z_{22}$=0).\cite{Goodenough63}
Hence, based on the experimentally determined $\theta_{c}$\,=\,80\,K for $B||c$ and the evaluated $J_2$\,$\sim$\,$-1.6J_1$ ratio, we estimate $J_1$\,$\approx$\,67\,K and $J_2$\,$\approx$\,$-107$\,K. Similarly, we obtain $J_1$\,$\approx$\,50\,K and $J_2$\,$\approx$\,$-80$\,K from $\theta_{a,b}$\,=\,60\,K for $B||a,b$. 
The large AFM $J_2$ value is in agreement with AFM component remaining beyond 8\,T, which reflects in the weak field dependence of magnetization above $B_C$. On the other hand, the derived difference between the exchange interactions parallel and perpendicular to the $c$-axis implies the presence of sizable ($\sim$\,20\,\%) exchange anisotropy. In fact, the difference between the main exchange interactions, $\approx$\,17\,K (i.e., $\sim$\,12.6\,T), is of the same order of magnitude as the observed and estimated saturation magnetic fields of $\sim$\,7 and $\sim$\,15\,T for $B||a$ and $B||b$, respectively.
The above agreement implies that the magnetic fields, needed to bend the magnetic moments out of the preferred orientation, indeed compensate the exchange anisotropy and is thus in line with the fact that zero-field splitting, imposed by crystal filed, for $S$\,=\,1/2 systems is negligible and with the apparent $g$-factor anisotropy, \cite{Bencini} reflected in the magnetic susceptibility. Still, the estimated anisotropy is relatively small ($\sim$\,20\,\%) compared to the magnitude of the exchange parameters, which implies that the dominant exchange is Heisenberg-like with a sizable exchange anisotropy.
The origin of this anisotropy might be the 2D nature of the magnetic lattice and the specific arrangement of the Cu$^{2+}$ atomic orbitals combined with spin-orbit and Coulomb exchange interactions.\cite{Yildirim95}

Finally, an estimate of the AFM interlayer exchange coupling $J'$ can be obtained, within the Weiss field model, from the magnitude of the magnetic field $B_C$\,=\,0.8\,T required to flip the layers and thus overcome $J'$: \cite{Goodenough63}
 \begin{equation}
 g \mu_B B_C= 2 z |J'| S.
 \end{equation}
Taking $z$\,=\,2, i.e., considering two neighboring layeres, the resulting $J'$ equals 0.5\,K.
Still, the significantly suppressed $T_N$ compared to $J_i$'s is not only a result of weak interlayer couplings, but probably also reflects the competition between AFM and FM interactions within the layer.

\section{CONCLUSIONS}

The anisotropic magnetic properties of the layered Kagom\'{e}-like system {\CBSOB} have been studied by bulk magnetic measurements and neutron diffraction.
We have found that below $T_N$\,=\,27.4\,K 
the magnetic ground state can be described as antiferromagnetically coupled $ab$ layers, with moments on the Cu2 sites pointing along $c$ and those on the Cu1 sites alternating between the $\pm50^\circ$ tilt from $c$ towards $b$.
At $B_C$\,$\sim$\,0.8\,T applied perpendicular to the layers (along $c$), every second layer flips, resembling a metamagnetic transition to an almost FM structure.

Based on the determined spin arrangements and critical fields we provide an estimate of the characteristic exchange couplings of the system. For intralayer exchange (along the $c$ axis) we thus obtain that nearest-neighbor Cu-O-Cu exchange is FM ($J_1$\,$\approx$\,$J_3$\,$\approx$\,67\,K), while next-nearest-neighbor Cu-O-Bi-O-Cu exchange is even stronger and AFM ($J_2$\,$\approx$\,--107\,K). 
Obviously, the nearest-neighbor FM interactions remove geometrical frustration, which would be a dominant feature of the Kagom\'{e} antiferromagnet; the frustration is only partially restored by next-nearest-neighbor AFM interactions, responsible for the canting of Cu1 moments. 
We note that considerably (20\,\%) lower $J_i$ values were derived for the perpendicular orientation (along the layers), implying a sizable exchange anisotropy.
Interestingly, in spite of the fact that the ratio between the interlayer and intralayer couplings is significant,  $|J'/J|$\,$\sim$\,0.006, and that the exchange interactions are predominantly isotropic, the intensity of the magnetic diffraction peak can be in a broad region below $T_N$ described as $(T-T_N)^{2\beta}$, with $\beta$\,$\leq$\,0.23, characteristic for finite-sized 2D XY magnetic systems with additional weak in-plane crystal-field anisotropy. This may imply that interlayer interactions are still weak enough that spin fluctuations below 0.96\,$T_N$ are essentially 2D, and that the existing magnetic anisotropy could be strong enough to impose a quasi 2D XY behavior.

\bigskip

{\bf Acknowledgements}\\
\\
This work was supported by the Swiss National Foundation (SNF) project 200021-129899, the Deutsche Forschungsgemeinschaft (DFG) via research unit 960 "Quantum Phase
Transitions" and the Transregional Collaborative Research Center TRR80 "From Electronic Correlations to Functionality" (Augsburg--Munich). We thank Dana Vieweg, Thomas Wiedenmann and Lukas Keller for experimental support. The neutron diffraction work was performed at SINQ, Paul Scherrer Institute, Villigen, Switzerland.


\begin{thebibliography}{99}
\bibitem{LowSim} We note that weak $hk$0 reflections with $h+k$\,=\,2$n$+1 are present up to room temperature, indicating that the actual crystal symmetry is lower than the proposed $Pmmn$.\cite{Millet01}
\end{thebibliography}
\end{document}